\documentclass[prl,aps,showpacs,twocolumn]{revtex4}
\usepackage{amsmath,amsfonts,bm}
\usepackage[dvips]{graphicx}
\usepackage{times}

\begin{document}

\title{Statistics of Entropy Production in Linearized
Stochastic System}

\author{K.
Turitsyn $^{a,b}$, M. Chertkov $^b$, V.Y. Chernyak $^{b,c}$,  A. Puliafito $^{b,d}$,  }

\affiliation{ $^a$ Landau Institute for Theoretical Physics, Moscow,
Kosygina 2, 119334, Russia\\
$^b$  Theoretical Division and Center for Nonlinear Studies, LANL, Los Alamos, NM 87545, USA\\
$^c$ Department of Chemistry, Wayne State University,
5101 Cass Ave, Detroit, MI 48202, USA\\
$^d$ INLN, 1361 route des Lucioles, F-06560, Valbonne, France
 }

\date{\today}

\begin{abstract}
We consider a wide class of linear stochastic problems driven off the equilibrium by a
multiplicative asymmetric force. The force brakes detailed balance, maintained otherwise, thus
producing entropy. The large deviation function of the entropy production in the system is
calculated explicitly. The general result is illustrated using an example of a polymer immersed in
a gradient flow and subject to thermal fluctuations.
\end{abstract}


\maketitle

The Gibbs distribution, ${\cal P}_{eq}({\bm x})\sim \exp[-U({\bm
x})/T]$, describes the probability for a system characterized by the
microscopic potential $U({\bm x})$ and maintained at equilibrium at
temperature $T$ to be observed in the state ${\bm x}$. In
particular, in our model case of a polymer the elastic potential
$U({\bm x})$ depends on the end-to-end position vector ${\bm x}$.
The system at equilibrium maintains the detailed balance, which is
the most fundamental principle of equilibrium statistical mechanics
\cite{24Tol,28Bri,28Nyq,foot1,99Cro}. Formally, the detailed balance
means that the probabilities ${\cal P}\{x\}$ and ${\cal P}\{x^*\}$
for a stochastic trajectory, $\{{\bm x}\}\equiv\{{\bm
x}(t');0<t'<t\}$ and its ``conjugated twin" $\{{\bm
x}\}^*\equiv\{{\bm x}^*(t')={\bm x}(t-t');0<t'<t\}$ are related by
\begin{equation}
\underline{\mbox{in balance:}}\quad\quad \ln\frac{{\cal
P}\{x\}}{{\cal P}\{x^*\}}=\frac{U({\bm x}(t))-U({\bm
x}(0))}{T}.\label{equi}
\end{equation}
Asymmetric external force breaks down the detailed balance. For example, a shearing flow forces the
polymer to tumble and results in steady entropy production \cite{77Hin,05CKLT}. In general,
configurational entropy is naturally defined as a mismatch between the left and right hand sides of
Eq.~(\ref{equi}):
\begin{equation}
 \underline{\small \mbox{off ballance:}}\ {\cal S}=
 \ln\frac{{\cal P}\{{\bm x}^*\}}{{\cal P}\{{\bm x}\}}
 +\frac{U({\bm x}(t))\!-\!U({\bm x}(0))}{T}\neq 0.
\label{entr}
\end{equation}
For a wide class of thermalized systems, driven out of equilibrium by external non-conservative
forces the entropy has also a standard thermodynamic interpretation: It determines the total heat
produced by the system over time $t$. In the off-detailed balance case entropy is a fluctuating
function of the entire configurational trajectory $\{{\bm x}\}$. Therefore, in the statistically
steady non-equilibrium case fluctuations occur on the top of a steady mean growth of the entropy
and one can argue that at sufficiently large observation time the distribution function of the
produced entropy ${\cal S}$ takes a large deviation form \cite{85Ell,99LS}:
\begin{eqnarray}
{\cal P}({\cal S}|t)\sim \exp\left[-t{\cal L}({\cal
S}\tau/t)/\tau\right],\label{LDF}
\end{eqnarray}
where $\tau$ is the typical correlation (turnover) time of the
system and ${\cal L}(\omega)$ is referred to as the large deviation
function. {\em Description of the large deviation function for a
truly non-equilibrium problem is a difficult task}, and only a few
successful results have been reported so far \cite{05Der}. To
clarify the difficulty let us also mention that a simpler problem of
finding an off-detailed balance analog of the Gibbs distribution,
posed in a classical work of Onsager \cite{31Ons}, has been solved
for a few examples only. ( See \cite{92Kam,04TK,05KAT} for
discussion of some difficulties, progress and results achieved on
this thorny path.) It is worth to note that the large deviation
function of entropy production was computed and verified
experimentally for a number of other physical situation, e.g.
optically dragged Brownian particles, electrical circuits and forced
harmonic oscillators \cite{03ZC}. Although these works are
ideologically similar, technically they study different
non-equilibrium systems, which are either non-steady or do not have
the detailed balance broken.

In this letter we present a {\em solution of this challenging task}
for a wide class of linear problems driven by multiplicative
asymmetric "force" and also connected to a Langevin reservoir. Such
problems arise whenever a statistically steady non-equilibrium state
is externally driven by space (${\bm x}$)-dependent non-conservative
external forces. In this context our main physical example is of a
Hookean polymer stretched and sheared by a mild constant external
flow. For this linear non-equilibrium setting we report  an explicit
expression for the large deviation function of the entropy
production.  For the most general case our result is given as a
solution of a well-defined system of algebraic equations or,
alternatively, in terms of an one-dimensional integral
Eq.~(\ref{Zint}). For the linear polymer in a constant gradient flow
the large deviation function is presented in terms of elementary
functions, see e.g. Eq.~(\ref{La}). The most important features of
the results for the large
deviation function derived in this Letter are:\\

\noindent$\bullet {\bf 1}$ The steady state solutions are consistent
with the fluctuation theorem \cite{93ECM,95GC,98Kur}:
 \begin{equation}
 {\cal L}(\omega)-{\cal L}(-\omega)=-\omega.\label{FT}
 \end{equation}

\noindent$\bullet {\bf 2}$ The large deviation function is found to
be very different from the Gaussian shape. Extreme tails of the
entropy PDF are exponential, that are the steepest tails allowed by
the large deviation form (\ref{LDF}).

\noindent$\bullet {\bf 3}$ One observes reduction in the number of
parameters affecting the shape of the large deviation function. For
example, the large deviation function is completely insensitive to
the symmetric part of the velocity gradient in the case of the
linear polymer immersed in a $2d$ flow.

The letter is organized as follows. After introducing the polymer in a flow example, we focus on
this model calculating the entropy production and defining the corresponding generating function.
We further derive a Fokker-Planck equation for the generating function and solve it using a
Gaussian ansatz for the related eigenvalue problem. This results in a system of matrix algebraic
equations that define the principle part of the longest-time asymptotic of the generating function.
The solution for the most general case of two-dimensional flow is expressed in terms of elementary
functions. This explicit description is also extended to a special three dimensional case. Then, we
switch to a general linear model with a constant multiplicative force. The general model, defined
in Eq.~(\ref{LinGen}), is analyzed using a direct solution of the linear stochastic equations
followed by averaging the resulting expression for the generating function over the Langevin noise.
This method is complementary to the aforementioned Fokker-Planck approach. The final result for the
large time asymtptotic of the generating function is presented in terms of the one-dimensional
integral over frequency.

The polymer's end-to-end vector, $x_i$ satisfies a stochastic equation of motion \cite{87BCAH}
\begin{eqnarray}
\dot{x}_i-\sigma_{ij}x_j=-\partial_{x_i}U({\bm x})+\xi_i,
\label{eq_mot}
\end{eqnarray}
where the traceless matrix $\hat{\sigma}$ describes the local value of the velocity gradient in the
generic incompressible flow ($i=1,\cdots,d$). The smoothness of velocity on the scale of an even
very extended polymer is justified by many experimental observations,  see e.g. \cite{87BCAH}.
Eq.(\ref{eq_mot}) describes the balance of forces: the second term on the lhs represents the
polymer deformation by the velocity field. The two terms on the rhs of Eq.(\ref{eq_mot}) account
for the polymer elasticity and for the Langevin thermal noise. In this Letter we consider a
deterministic constant gradient flow, $\hat{\sigma}=\mbox{const}$ and focus on the case of a
relatively weak $\hat{\sigma}$ and linear model of Hookean polymer, $U_H({\bm x})={\bm
x}^2/(2\tau)$. Yet, we will also discuss the other extreme of a stretched nonlinear polymer in a
strong gradient flow. The Langevin term in Eq.(\ref{eq_mot}) is modeled by the zero mean white
Gaussian noise with $\langle \xi_i(t)\xi_j(t')\rangle=2T\delta(t-t')\delta_{ij}$.

According to Eq.~(\ref{eq_mot}) the probability for a stochastic polymer trajectory in the
configuration space is given by
\begin{eqnarray}
 {\cal P}\{{\bm x}\}\sim \exp\left(-\int_0^tdt'
 \left(\dot{\bm x}-\hat{\sigma}{\bm x}+\partial_{\bm
 x}U({\bm x})\right)^2/(4T)\right),
 \nonumber
\end{eqnarray}
with the entropy production (\ref{entr}) along the path $\{{\bm x}\}$ (see e.g. \cite{06CCJ} for
details)
\begin{eqnarray}
 {\cal S}=\int_0^tdt' \dot{x}^\alpha(t')\left(\sigma^{\alpha\beta}-
 \sigma^{\beta\alpha}\right)x^\beta(t')/(2T)+{\cal O}(1), \label{Entr}
\end{eqnarray}
where the longest-time $t/\tau\gg 1$ statistics of ${\cal S}$ is
naturally described within ${\cal O}(t/\tau)$.

The Fokker-Planck technique is applied via relating the large deviation function to its generating
function defined by
\begin{eqnarray}
Z_q\equiv \langle\exp\left(-q{\cal S}\right)\rangle_{{\bm \xi}}.
\label{Qn}
\end{eqnarray}
Utilizing the large deviation asymptotic (\ref{LDF}) we can
reformulate averaging over the Langevin noise in Eq.~(\ref{Qn}) in
terms of integration over ${\cal S}$ and further evaluate the
integral using the saddle-point approximation (justified for the
asymptotically long time). This establishes the following Legendre
transform relation between the generating and the large deviation
functions:
\begin{eqnarray}
Z_q\sim \exp\left(-\left({\cal L}(\omega_q)-\omega_q{\cal
L}'(\omega_q)\right)t/\tau\right),\quad -q={\cal L}'(\omega_q).
\label{Qn-Cra}
\end{eqnarray}
Substituting Eqs.~(\ref{Entr}) into the definition (\ref{Qn}) followed by averaging over the
stochastic dynamics (\ref{eq_mot}) we arrive at the Fokker-Planck equation:
\begin{eqnarray} &&
\partial_t Z_q=\hat{L}_q Z_q,
\label{SL} \\
&&
\hat{L}_q=-\nabla^\alpha\left(\sigma^{\alpha\beta}x^\beta-\partial_{x^\alpha}
U({\bf x})\right)+T\nabla^\alpha \nabla^\alpha,\label{Ham}\\ &&
\nabla^\alpha=\partial^\alpha+
q(\hat{\sigma}-\hat{\sigma}^+)^{\alpha\beta}x^\beta/(2T), \nonumber
\end{eqnarray}
where the ``gauge" term ``elongates" the derivatives in a standard
way. Assuming that the operator $\hat{L}_q$ has a discrete spectrum,
the large time asymptotics of $Z_q$ is completely dominated by the
ground state of $\hat{L}_q$. Therefore, the large deviation function
is fully described by its ground-state eigen-value and,
specifically, its dependence on $q$.

Here we consider the case when velocity gradient is relatively weak compared to the elastic force
(or when stretching components in $\hat{\sigma}$ are zero) so that the steady state is achieved in
the regime when elasticity in Eq.(\ref{Ham}) is linear, $U({\bf x})={\bm x}^2/(2\tau)$ (the
so-called coiled, rather then stretched, state.) The spectral problem (\ref{SL}) in the case of
Hookean elasticity is of an "integrable" type and is similar to a single-particle quantum mechanics
in a constant magnetic field, when the ground-state eigen-function is Gaussian. Therefore, one
looks for the solution of Eq.~(\ref{SL}) in a form $Z_q=\exp(-\lambda_q t)\exp[-x_i B^{ij}_q
x_j/(2T)]$, with $\hat{B}$ being a $d\times d$ symmetric matrix. Substituting the Gaussian ansatz
into Eq.~(\ref{SL}) we derive the following eigen-value relations for $\hat{B}_q$ and $\lambda_q$:
\begin{eqnarray}
 &&\lambda_q=tr\left(\hat{B}_q+\hat{\sigma}-\hat{1}/\tau\right),\quad
\hat{M}+\hat{M}^+=0, \label{lam2}\\
&&
 \hat{M}\!\equiv\!\left(\!\hat{B}_q\!+\!q(\hat{\sigma}\!-\!\hat{\sigma}^+)/2\right)
 \!\!\left(\hat{\sigma}\!-\!\hat{1}/\tau\!+\!\hat{B}_q\!-\!
 q(\hat{\sigma}\!-\!\hat{\sigma}^+)/2\right),
 \nonumber
\end{eqnarray}
that correspond to $x^2\exp(...)$ and $\exp(...)$ terms, respectively.
 Eqs.~(\ref{lam2}) are generic,  i.e. valid for the full
formulation given by Eqs.(\ref{SL},\ref{Ham}); the $\lambda_0=0$ and the $q=0$ version of
Eqs.~(\ref{lam2}) defines the steady distribution function for the end-to-end polymer length ${\bm
x}$ (see, e.g. \cite{05KAT}).

\begin{figure}[tl]
\includegraphics[width=0.38\textwidth]{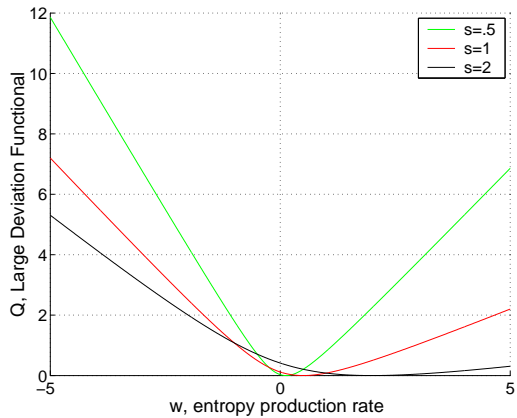}
\caption{Large deviation function in $d=2$ for three values of the
governing parameter $c\tau=1.;2.;4.$ (red;green;blue).} \label{Q_w}
\end{figure}

For the $2\times 2$ traceless velocity gradient matrix
$\sigma_{11}=-\sigma_{22}=a$ and $\sigma_{12}=b+c$,
$\sigma_{21}=b-c$ one derives from Eq.~(\ref{lam2})
\begin{equation}
\lambda_q=\left(\sqrt{1+4q(1-q)c^2\tau^2}-1\right)\tau^{-1},
 \label{lamb}
\end{equation}
where the generating function  and the eigenvalue of the ground
state are well defined only within a finite interval, $q\in
[q_-;q_+]$, where $q_\pm\equiv 1/2\pm\sqrt{1/4-1/(c^2\tau^2)}$.
Notice that the eigen-value $\lambda_q$ given by Eq.~(\ref{lamb})
does not depend on the symmetric part of $\hat{\sigma}$,  even
though the resulting $\hat{B}_q$ function does depend explicitly on
both symmetric and anti-symmetric components. Formally this reflects
the invariance of the Fokker-Planck operator with respect to a
family of iso-spectral transformations that keep the spectra (or at
least its ground state) invariant. We do not know however a
physically intuitive explanation for this remarkable
symmetry/reduction in the model. A similar and equally surprising
phenomenon of reduction in the degrees-of-freedom number that
control the large deviation function of a current has been recently
reported for a different non-equilibrium system that models a
contact between two thermostats kept at different temperatures
\cite{05Der}. Combining Eqs.~(\ref{Qn-Cra},\ref{lamb}) leads to
\begin{equation}
{\cal L}(\omega)=
 \sqrt{(1+c^2\tau^2)(4c^2\tau^2+\omega^2)}-1-\omega/2.\label{La}
\end{equation}
Note that, first, ${\cal L}(\omega)$ satisfies the fluctuation
theorem (\ref{FT}); and, second, the asymptotics of ${\cal
L}(\omega)$ at $|\omega|\gg c\tau$ are both linear in $\omega$,
making the extreme deviation asymptotics of the entire distribution
function of ${\cal S}$ exponential and time independent. Tracking
the origin of the exponential tails back to a special form of the
generating function (\ref{lamb}), one finds that these extreme
asymptotics correspond to the square root singularity of $\lambda_q$
at $q_\pm$. Time-independence of the ${\cal P}(S|t)$ asymptotics
means that typical trajectories contributing the PDF tail are
correlated at some finite times, so that the observational time $t$
increase does not change corresponding probabilities. Figure
\ref{Q_w} shows the large deviation function given by Eq.~(\ref{La})
verified versus Brownian dynamics simulations. In the case of a
$d=3$ gradient flow the algebraic system of Eqs.~(\ref{lam2}) is too
complicated to allow  a solution in terms of elementary functions
for an arbitrary form of the velocity gradient matrix
$\hat{\sigma}$. Thus we mention only a special example of a $3d$
flow with the following nonzero elements of $\hat{\sigma}$:
$\sigma_{11}=-a_1, \sigma_{22}=a_1+a_2, \sigma_{33}=-a_2,
\sigma_{13}=c, \sigma_{31}=-c$. In this case the generating function
and the large deviation function is given by Eqs.~(\ref{La})
modified according to the following simple renormalization of
$\tau\to \tau/|1+a\tau|$ (and $\omega$ respectively).

An alternative derivation of the large deviation function starts with considering a general linear
problem
\begin{equation}
\dot{x}_i=\Phi_{ij}x_j+\Upsilon_{ij}\xi_j, \label{LinGen}
\end{equation}
where $\hat{\Phi}$ and $\hat{\Upsilon}$ are arbitrary constant matrices. Eq.~(\ref{LinGen})
describes linear stochastic dynamics around a fixed point,  that can be stable or unstable. We
discuss here a truly non-equilibrium (off-detailed balance) steady state maintained if the
fluctuations do not exceed a threshold so that nonlinear effects can be ignored (see \cite{05KAT}
and references therein). A flux state observed in diffusive system \cite{01DL} is a popular example
that involves an infinite-dimensional configurational variable ${\bm x}$. Many examples of the
off-detailed balance steady systems, e.g. vesicles or red-blood cells  in external flows
\cite{05KS} and macromolecular biological devices, such as enzyme motors \cite{03OW}, come from
biology and soft-matter physics. Obviously, the $\hat{\Upsilon}=\hat{1}$ version of
Eq.~(\ref{LinGen}) describes the aforemention two-beads Hookean polymer model as well, however it
is worth mentioning that the full version of Eq.~(\ref{LinGen}) also appears naturally in a more
general polymer context, where the $\hat{\Upsilon}\neq\hat{1}$ case models hydrodynamic
interactions between the different parts of the polymer chain \cite{87BCAH}. Eq.~(\ref{LinGen})
also describes fluctuations around a stretched state above the coil-stretch transition
\cite{73Lum,00BFL,00Che} in a strong gradient flow \cite{footnote}.

According to Eq.~(\ref{entr}) the entropy production in the system
described by Eq.~(\ref{LinGen}) is given by
 ${\cal S} = \int_0^t d t' \dot{{\bm x}}^+(t')
 (\hat{K}\hat{\Phi}-\hat{\Phi}^+ \hat{K}) {\bm x}(t')/(2T)$,
where $\hat{K}\equiv\left(\hat{\Upsilon}\hat{\Upsilon}^+\right)^{-1}$. To discuss dynamics at large
but finite time $t$  it is convenient to invoke a discrete (a-la Matsubara) frequency
representation
\begin{eqnarray}
 && {\bm x}(t') =\sum_k \left({\bm c}_k \exp(i \omega_k t')+{\bm c}_k^* \exp(-i \omega_k t')\right)/\sqrt{t},
 \nonumber\\
 && Z_q \!= \!\int \!\!\prod_k\!\! {\cal D} c_k{\cal D} c_k^*
 \exp\!\left(\!- \sum_k c_k^\dag A_q(\omega_k) c_k/(2T)\!\right), \label{Mats2} \\
 && A_q(\omega)\equiv \omega^2 \hat{K} + \hat{\Phi}^+ \hat{K}\hat{\Phi} +
 i \omega(1-2 q) (\hat{K}\hat{\Phi}-\hat{\Phi}^+ \hat{K}), \nonumber
\end{eqnarray}
where  $\omega_k\equiv 2\pi k/t$, $k=1,2,\dots$. Straightforward
Gaussian integration in Eq.~(\ref{Mats2}) yields: $\lambda_q t
=\sum_k \log(\det A_q(\omega_k)/\det A_0(\omega_k))$. In the
$t\to\infty$ limit one replaces summation over $k$ by integration
and arrives at
\begin{eqnarray}
 \lambda_q=\int_{0}^{\infty} \frac{d\omega}{2\pi} \log\left(
 \frac{\det A_q(\omega)}{\det A_0(\omega)}\right), \label{Zint}
\end{eqnarray}
which is the most general long-time asymptotic result reported in
this letter. It is straightforward to verify that in the special
case of linear polymer advected by $d=2$ gradient flow,
$\hat{\Phi}=\hat{\sigma}-\hat{1}/\tau$, and Langevin driving,
$\hat{\Upsilon}=\hat{1}$, the integral representation (\ref{Zint})
for $\lambda_q$ turns into Eq.~(\ref{lamb}) derived earlier using
the spectral method.

Note that Eq.~(\ref{Zint}) also suggests a convenient way to
determine the values of $q_{\pm}$ for a general linear system
(\ref{LinGen}). One finds that $q_+$ is a minimal positive value of
$q$ for which a solution of the equation $\det A_q(\omega)  = 0$
does exist. The value of $\omega=\omega_*$ which solves this
equation is related to the characteristic time-scale of an optimal
fluctuation which controls the exponential tail of the entropy PDF.
For the two-dimensional polymer problem this leads to the following
explicit expression for $\omega_* =\sqrt{1+c^2- a^2 - b^2}/\tau$.
The limit of $\omega_* \to 0$ corresponds to the coil-stretch
transition.

We conclude compiling an incomplete list of future challenges related to the approach and results
reported in this Letter. First, our analysis of the entropy production in a polymer system extends
to the case of chaotic flows, e.g. realized in the recently discovered elastic turbulence
\cite{00GS}. Of a particular interest here is to check sensitivity of the large deviation function
to the coil-stretch transition observed in the chaotic problem \cite{73Lum,00BFL,00Che}. Second,
introducing a finite time protocol for a controlled parameter (e.g. shear in the polymer solution
experiments) one may be interested to go beyond the analysis of the stationary problem, in
particular discussing an off-detailed balance version of the Jarzynski relation
\cite{97Jar,99Cro,99Hat,05CCJ,06CCJ}. This may also help to reconstruct (experimentally or
numerically) the non-equilibrium steady-state landscape that would be akin to experiments currently
developed for the systems that do not violate the detailed balance, see e.g. \cite{05CRJSTB}. Newly
developed experimental techniques that allow to track polymers in regular \cite{03Chu} and chaotic
\cite{05GCS} flows may be very useful for such applications. Finally, entropy production is closely
related to the flux over spatial scales in the Kolmogorov-like cascade solutions of turbulence
\cite{95Fri}. Studying fluctuations of the entropy production in turbulence by experimental,
numerical and theoretical means may have a tremendous impact on understanding of this most
challenging problem in non-equilibrium physics.

This work was carried out under the auspices of the National Nuclear
Security Administration of the U.S. Department of Energy at Los
Alamos National Laboratory under Contract No. DE-AC52-06NA25396. VYC
also acknowledges the support through WSU. KT acknowledges the support
by ENS and INTAS.


\begin{thebibliography}{99}

\bibitem{24Tol} R.C.~Tolman, Phys.Rev. {\bf 23}, 693 (1924).

\bibitem{28Bri} P.W.~Bridgman, Phys.Rev. {\bf 31}, 90 (1928).

\bibitem{28Nyq} H. Nyqist, Phys.Rev. {\bf 32}, 110 (1928).

\bibitem{foot1} Strictly speaking Eq.~(\ref{equi}) should be called ``miscroscopic
reversibility" while the ``detailed balance" is referred to a
probability of changing states without a reference to a particular
temporal path. See e.g. \cite{99Cro} for modern discussion of the
terminological nuance.

\bibitem{99Cro} G.E. Crooks,
Phys.Rev.E. {\bf 60}, 2721-2726 (1999).

\bibitem{77Hin}
 E.~J.~Hinch, Phys. Fluids, {\bf 20}, S22 (1977).

\bibitem{05CKLT} M. Chertkov, I. Kolokolov, V. Lebedev and K. Turitsyn,
Journal of Fluid Mechanics {\bf 531}, 251-260 (2005).


\bibitem{85Ell} R. Ellis, Entropy, Large Deviations and Statistical Mechanics,
 Springer-Verlag, Berlin, 1985.

\bibitem{99LS} J.L. Lebowitz, H. Spohn,
J. Stat.Phys. {\bf 95}, 333 (1999).

\bibitem{05Der} B.~Derrida,
Pramana J.of Physics {\bf 64}, 695
(2005).

\bibitem{31Ons} L. Onsager, Phys.Rev. {\bf 37}, 405 (1931).

\bibitem{92Kam} N.G. van Kampen, {\it Stochastic Processes in Physics and
Chemistry}, Elsevier, Amsterdam, 1992.


\bibitem{04TK} S. Tanase-Nicola, J. Kurchan, J.Stat.Phys. {\bf 116},
1201 (2004).

\bibitem{05KAT} C. Kwon, P. Ao, D.J. Thouless,
PNAS {\bf 102}, 13029 (2005).

\bibitem{03ZC} R. Van Zon, E. G. D. Cohen, Phys. Rev. E. {\bf 67}, 461021 (2003);
R. Van Zon, S. Ciliberto, E. G. D. Cohen,Phys Rev Lett {\bf 92}, 130601 (2004);
N. Garnier, S. Ciliberto, Phys. Rev. E,  {\bf 71}, 060101 (2005);
F. Douarche, S. Joubaud, N. B. Garnier, A. Petrosyan, S. Ciliberto, Phys Rev Lett, {\bf 97}, 140603 (2006).

\bibitem{93ECM} D.J. Evans, E.G.D. Cohen, G.P. Morris, Phys.Rev.Lett. {\bf 71}, 2401 
(1993).

\bibitem{95GC} G. Gallavotti, E.G.D. Cohen,
Phys.Rev.Lett. {\bf 74}, 2694 (1995).

\bibitem{98Kur} J. Kurchan,
J.Phys.A {\bf 31}, 3719 (1998).


\bibitem{87BCAH} R.B. Bird, C.F. Curtiss, R.C. Armstrong, O.
Hassager, {\it Dynamics of polymeric liquids}, vol.2, Wiley \& Sons,
1987.

\bibitem{06CCJ} V.~Chernyak, M.~Chertkov and C.~Jarzynski,
J.~Stat.~Mech. P08001 (2006).

\bibitem{00GS}
 A. Groisman and V. Steinberg ,
 Nature {\bf 405}, 53 (2000).

\bibitem{01DL} B. Derrida, J. L. Lebowitz, E. R. Speer,  Phys Rev Lett {\bf 87}, (2001);
T. Bodineau, B. Derrida, Phys. Rev. E {\bf 72},066110 (2005).
\bibitem{05KS} V. Kantsler, V. Steinberg, Phys Rev Lett {\bf 95}, 258101 (2005);
M. Kraus, W. Wintz, U. Seifert, R. Lipowsky, Phys Rev Lett {\bf 77},
3685 (1996).

\bibitem{03OW} G. Oster, H. Wang, Trends Cell Biol. {\bf 13}, 114 (2003);
U. Seifert, Europhys. Lett. {\bf 70}, 36 (2005).


\bibitem{73Lum}
 J.~L.~Lumley,
 Annu. Rev. Fluid Mech. {\bf 1}, 367 (1969);
 J. Polymer Sci.: Macromolecular Reviews {\bf 7}, 263 (1973).

\bibitem{00BFL} E. Balkovsky, A. Fouxon, V. Lebedev,
Phys. Rev. Lett. {\bf 84}, 4765 (2000); Phys. Rev. E {\bf 64},
056301 (2001).

\bibitem{00Che} M. Chertkov,Phys. Rev. Lett. 84, 4761 (2000).

\bibitem{footnote} Nonlinear polymer in a strong gradient flow is described by
Eq.~(\ref{eq_mot}) with $U({\bm x}) = \gamma(x) {\bm x}^2/2$, where the nonlinear relaxation rate,
$\gamma(x)$, is a monotonic function of the polymer length with $\gamma(0) =
\tau^{-1}<\gamma(x_{max}) = +\infty$, and the largest eigenvalue $\mu$ of $\hat{\sigma}$ satisfies
$\mu \tau
> 1$ \cite{00BFL,00Che}. Deterministic polymer dynamics has a fixed point
${\bm X}$: $\hat{\sigma} {\bm X} = \gamma(X) {\bm X}$. Thermal fluctuations around the stable fixed
point, ${\bm y} = {\bm x}-{\bm X}$, are described  by the linear equation $\dot{{\bm y}} =
\hat{\sigma} {\bm y} - \gamma(X) { \bm y} - \gamma'(X) ({\bm X} {\bm y}) {\bm X} +{\bm \xi} $, that
is a particular case of Eq.~(\ref{LinGen}). The equation is valid only for small fluctuations near
the fixed points ${\bf X}$ or $-{\bf X}$, present in the problem. It does not explain thermally
driven tunneling (tumbling) between the fixed points. However, for syfficiently small temperature
the tunneling is weak and its contribution to the total entropy production is negligible.

\bibitem{97Jar} C. Jarzynski,
Phys. Rev. Lett. {\bf 78}, 2690 (1997); Phys.Rev. E {\bf 56}, 5018
(1997).

\bibitem{99Hat} T. Hatano, Phys.Rev.E. {\bf 60}, R5017 (1999).


\bibitem{05CCJ} V.~Chernyak, M.~Chertkov and C.~Jarzynski,
Phys.Rev. E {\bf 71}, 025102 (2005).

\bibitem{05CRJSTB}
D.~Collin, F.~Ritort, C.~Jarzynski, S.B.~Smith, I.~Tinoco,
C.~Bustamante, Nature {\bf 437}, 231 (2005).

\bibitem{03Chu} S. Chu, Phil. Trans. R. Soc. Lond. A 361, 689
(2003).

\bibitem{05GCS} S. Gerashchenko, C. Chevallard, and V. Steinberg,
Europhys.Lett.  {\bf 71} 221 (2005).


\bibitem{95Fri} U.~Frisch, {\it Turbulence. The legacy of A.N.
Kolmogorov}, Cambridge U. Press, 1995.

\end{thebibliography}
\end{document}